\documentclass{ws-ijmpe}
\begin{document}

\markboth{G. Saxena, D. Singh, M. Kaushik, H. L. Yadav and H.
Toki}{Study of Neutron Magic Drip-Line Nuclei within RMF+BCS
Approach}

\catchline{}{}{}{}{}
%%%%%%%%%%%%%%%%%%%%%%%%%%%%%%%%%%%%%%%%%%%%%%%%%%%%%%%%%%%%%%%%%%%%

\title{STUDY OF NEUTRON MAGIC DRIP-LINE NUCLEI WITHIN RELATIVISTIC MEAN-FIELD PLUS BCS APPROACH}

\author{\footnotesize G. SAXENA}

\address{Department of Physics, Govt. Women Engineering College, Ajmer-305002, India
\footnote{Corresponding author}\\
gauravphy@gmail.com}

\author{D. SINGH}

\address{Department of Physics, University of Rajasthan, Jaipur-302004,
India}
\author{M. KAUSHIK}
\address{Department of Physics, University of Rajasthan,
Jaipur-302004, India}
\author{H. L. YADAV}
\address{Department of Physics, Banaras Hindu University,
Varanasi-221005, India\\ hlyadavphysics@gmail.com}
\author{H. TOKI}
\address{Research Center for Nuclear Physics (RCNP), Osaka University, 10-1,
Mihogaoka, Ibaraki, Osaka 567-0047, Japan}

\maketitle

\begin{history}
\received{Day Month Year}
\revised{Day Month Year}
%\accepted{Day Month Year}
%\comby{(xxxxxxxxxx)}
\end{history}

\begin{abstract}
Encouraged by the success of RMF+BCS approach for the description of
the ground state properties of the chains of isotopes of proton
magic nuclei with proton number \textit{Z}=8, 20, 28, 50 and 82 as
well as those of proton sub-magic nuclei with \textit{Z}=40, we have
further employed it, in an analogous manner, for a detailed
calculations of the ground state properties of the neutron magic
isotones with neutron number \textit{N} = 8, 20, 28, 50, 82 and 126
as well as those of neutron sub-magic isotones with \textit{N} = 40
using the TMA force parameterizations in order to explore low lying
resonance and other exotic phenomenon near drip-lines. The results
of these calculations for wave-function, single particle pairing
gaps etc. are presented here to demonstrate the general validity of
our RMF+BCS approach. It is found that, in some of the proton-rich
nuclei in the vicinity of the proton drip-line, the main
contribution to the pairing correlations is provided by the
low-lying resonant states, in addition to the contributions coming
from the states close to the Fermi surface, which results extended
proton drip-line for isotonic chain.
\end{abstract}


\ccode{21.10.-k, 21.10.Ft, 21.10.Dr, 21.10.Gv, 21.10.-n, 21.60.Jz}

%\tableofcontents

\section{Introduction}

During the last few years, phenomenal advances in nuclear
experimental techniques and the spectacular growth in accelerator
and detection technology have led to the development of radioactive
nuclear beams facilities around the world
\cite{tanihata,kanungo1,ozawa,ozawa1,hinkle,kanungo,tanaka}. With
these remarkable efforts it has been possible to study exotic nuclei
in the region far from the line of $\beta$-stability in the nuclear
chart. As far as theoretical studies are concern for these nuclei in
neutron- and proton-rich side they have been mostly carried out
within the framework of mean-field theories
\cite{dobac,dobac1,terasaki,grasso,sand,sand1}, as well as employing
their relativistic counterparts \cite{walecka}-\cite{yadav1}. The
effect of continuum on the pairing energy contribution within the
HF+BCS+Resonant continuum approach has been calculated by Grasso
{\it et al}. \cite{grasso} and Sandulescu {\it et al}.
\cite{sand,sand1}. Similarly, the effect of inclusion of positive
energy resonant states on the pairing correlations has been
investigated quite elaborately for the unstable nuclei by Yadav {\it
et al.} \cite{yadav,yadav1} within the framework of relativistic
mean-field (RMF) theory.

RMF approach has proved to be very crucial for the study of unstable
nuclei near the drip-line because it provides the spin-orbit
interaction in the entire mass region in a natural way
\cite{walecka,boguta,serot,pgr,pgr1}, and the single particle
properties near the threshold are prone to large changes as compared
to the case of deeply bound levels in the nuclear potential for
drip-line nuclei. Indeed the RMF+BCS scheme \cite{yadav1} yields
results which are in close agreement with the experimental data and
with those of recent continuum relativistic Hartree-Bogoliubov
(RCHB) and other similar mean-field calculations \cite{meng5,meng6}.

In the present investigations we have further employed the
relativistic mean-field plus BCS approach
\cite{yadav,yadav1,saxena,saxena1,geng1} to carry out a systematic
study for the ground state properties of the entire chains of
neutron magic nuclei represented by isotones of traditional neutron
magic numbers \textit{N} = 8, 20, 28, 50, 82 and 126 as well as
isotones of \textit{N} = 40, considered to be neutron sub-magic
\cite{broda95}. These chains of magic isotones cover the different
regions of the periodic table and their study is expected to provide
further testing ground for the general validity of the RMF+BCS
approach, including  for the magic isotones in the region away from
the line of $\beta$-stability up to the drip-lines. Additionally,
these calculations are expected to shed light on the evolution of
neutron and/or proton magicity as we move away from line of
$\beta$-stability to approach the proton or neutron drip-line. This
study of isotones will provide additional impetus on the recent
developments on isotones \cite{geng2,Li,sorlin,smirnova}.

\section{Relativistic Mean-Field Model}

In our RMF approach we have used the model Lagrangian density with
nonlinear terms both for the ${\sigma}$ and ${\omega}$ mesons. Our
calculations are performed with the TMA parametrization as described
in detail in Refs.~\refcite{suga,yadav,saxena,saxena1}.

\begin{eqnarray}
       {\cal L}& = &{\bar\psi} [\imath \gamma^{\mu}\partial_{\mu}
                  - M]\psi\nonumber\\
                  &&+ \frac{1}{2}\, \partial_{\mu}\sigma\partial^{\mu}\sigma
                - \frac{1}{2}m_{\sigma}^{2}\sigma^2- \frac{1}{3}g_{2}\sigma
                 ^{3} - \frac{1}{4}g_{3}\sigma^{4} -g_{\sigma}
                 {\bar\psi}  \sigma  \psi\nonumber\\
                &&-\frac{1}{4}H_{\mu \nu}H^{\mu \nu} + \frac{1}{2}m_{\omega}
                   ^{2}\omega_{\mu}\omega^{\mu} + \frac{1}{4} c_{3}
                  (\omega_{\mu} \omega^{\mu})^{2}
                   - g_{\omega}{\bar\psi} \gamma^{\mu}\psi
                  \omega_{\mu}\nonumber\\
               &&-\frac{1}{4}G_{\mu \nu}^{a}G^{a\mu \nu}
                  + \frac{1}{2}m_{\rho}
                  ^{2}\rho_{\mu}^{a}\rho^{a\mu}
                   - g_{\rho}{\bar\psi} \gamma_{\mu}\tau^{a}\psi
                  \rho^{\mu a}\nonumber\nonumber\\
                &&-\frac{1}{4}F_{\mu \nu}F^{\mu \nu}
                  - e{\bar\psi} \gamma_{\mu} \frac{(1-\tau_{3})}
                  {2} A^{\mu} \psi\,\,,%\nonumber\
\end{eqnarray}
where $H$, $G$ and $F$ are field tensors for the vector fields and
defined by
\begin{eqnarray}
                 H_{\mu \nu} &=& \partial_{\mu} \omega_{\nu} -
                       \partial_{\nu} \omega_{\mu}\nonumber\\
                 G_{\mu \nu}^{a} &=& \partial_{\mu} \rho_{\nu}^{a} -
                       \partial_{\nu} \rho_{\mu}^{a}
                     -2 g_{\rho}\,\epsilon^{abc} \rho_{\mu}^{b}
                    \rho_{\nu}^{c} \nonumber\\
                  F_{\mu \nu} &=& \partial_{\mu} A_{\nu} -
                       \partial_{\nu} A_{\mu}\,\,,\nonumber\
\end{eqnarray}
and other symbols have their usual meaning. We perform a state
dependent BCS calculations\cite{lane,ring2} on the basis of
single-particle spectrum calculated by the RMF described above. The
continuum is replaced by a set of positive energy states generated
by enclosing the nucleus in a spherical box. Thus the gap equations
have the standard form for all the single particle states, i.e.
\begin{eqnarray}
     \Delta_{j_1}& =&\,-\frac{1}{2}\frac{1}{\sqrt{2j_1+1}}
     \sum_{j_2}\frac{\left<{({j_1}^2)\,0^+\,|V|\,({j_2}^2)\,0^+}\right>}
      {\sqrt{\big(\varepsilon_{j_2}\,-\,\lambda \big)
       ^2\,+\,{\Delta_{j_2}^2}}}\,\,\sqrt{2j_2+1}\,\,\, \Delta_{j_2}\,\,,
\end{eqnarray}\\
where $\varepsilon_{j_2}$ are the single particle energies, and
$\lambda$ is the Fermi energy, whereas the particle number condition
is given by $\sum_j \, (2j+1) v^2_{j}\, = \,{\rm N}$. A delta force
i.e., V = -V$_0 \delta(r)$ is used in the calculations for the
pairing interaction, with the same strength V$_0$ for both protons
and neutrons. The value of this interaction strength V$_0$ = 350 MeV
fm$^3$ was determined in Ref.~\refcite{yadav} by obtaining a best
fit to the  binding energy of Ni isotopes. We use the same value of
V$_0$ for our present studies of isotopes of other nuclei as well.
Apart from its simplicity, the applicability and justification of
using such a $\delta$-function form of interaction has been
discussed in Refs.~\refcite{dobac,bertsch91,migdal67}. The pairing
matrix element for the $\delta$-function force is given by
\begin{eqnarray}
\left<{({j_1}^2)\,0^+\,|V|\,({j_2}^2)\,0^+}\right>&
=&\,-\,\frac{V_0}{8\pi}
       \sqrt{(2j_1+1)(2j_2+1)}\,\,I_R\,\,,
\end{eqnarray}
where $I_R$ is the radial integral with the form
\begin{eqnarray}
   I_R& =&\,\int\,dr \frac{1}{r^2}\,\left(G^\star_{j_ 1}\, G_{j_2}\,+\,
     F^\star_{j_ 1}\, F_{j_2}\right)^2
\end{eqnarray}
Here $G_{\alpha}$ and $F_{\alpha}$ denote the radial wave functions
for the upper and lower components, respectively, of the nucleon
wave function expressed as
\begin{equation}\psi_\alpha={1 \over r} \,\, \left({i \,\,\, G_\alpha \,\,\,
 {\mathcal Y}_{j_\alpha l_\alpha m_\alpha}
\atop{F_\alpha \, {\sigma} \cdot \hat{r}\, \, {\mathcal Y}_{j_\alpha
l_\alpha m_\alpha}}} \right)\,\,,
\end{equation}
and satisfy the normalization condition
 \begin{eqnarray}
         \int dr\, {\{|G_{\alpha}|^2\,+\,|F_{\alpha}|^2}\}\,=\,1
 \end{eqnarray}

In Eq. (5) the symbol ${\mathcal Y}_{jlm}$ has been used for the
standard spinor spherical harmonics with the phase $i^l$. The
coupled field equations obtained from the Lagrangian density in (1)
are finally reduced to a set of simple radial equations
\cite{pgr,pgr1} which are solved self consistently along with the
equations  for the state dependent pairing gap $\Delta_{j}$ and the
total particle
number \textit{N} for a given nucleus.\\

The relativistic mean-field description has been extended for the
deformed nuclei of axially symmetric shapes by Gambhir, Ring and
their collaborators \cite{gambhir,gambhir1} using an expansion
method. The treatment of pairing has been carried out in
Ref.~\refcite{geng1} using state dependent BCS method \cite{lane} as
has been given by Yadav et al. \cite{yadav,yadav1} for the spherical
case. For axially deformed nuclei the rotational symmetry is no more
valid and the total angular momentum $j$ is no longer a good quantum
number. Nevertheless, the various densities still are invariant with
respect to a rotation around the symmetry axis. Here we have
 taken the symmetry axis to be the \textit{z}-axis. Following Gambhir {\it
et al.} \cite{gambhir,gambhir1}, it is then convenient to employ the
cylindrical coordinates.

The scalar, vector, isovector and charge densities, as in the
spherical case, are expressed in terms of the spinor $\pi_i$, its
conjugate $\pi_i^+$, operator $\tau_3$ etc. These densities serve as
sources for the fields $\phi$ = $\sigma$, $\omega^0$ $\rho^0$ and
$A^0$, which are determined by the Klein-Gordon equation in
cylindrical coordinates. Thus a set of coupled equations, namely the
Dirac equation with potential terms for the nucleons and the
Klein-Gordon type equations with sources for the mesons and the
photon is obtained. These equations are solved self consistently.
For this purpose, as described above, the well-tested basis
expansion method \cite{vautherin73} has been employed
\cite{gambhir,gambhir1}. Using anisotropic (axially symmetric)
harmonic oscillator potential the bases are generated. These bases
are used to expand the upper and lower components of the nucleon
spinors, the fields, the baryon currents and densities separately.
In this expansion method the solution of the Klein-Gordon equation
reduces to a set of inhomogeneous equations while the Dirac equation
gets reduced to a symmetric matrix diagonalization problem. Solution
of these equations provide the spinor fields, and the nucleon
currents and densities (sources of the fields). From these fields
all the relevant ground state nuclear properties are calculated. For
further details of these formulations we refer the reader to
Ref.~\refcite{gambhir,gambhir1,saxena,geng1}.

\section{Results and Discussion}

\subsection{\textit{N} = 28 Isotones: A Representative Case of Neutron Magic
Nuclei}

It has been earlier demonstrated in the case of neutron-rich
isotopes of proton magic nuclei \cite{yadav1} that the neutron
resonant states lying near the Fermi level have their wave functions
mostly confined within the nuclear potential region. Consequently,
these resonant states have properties akin to a bound state. This
enables a resonant state to accommodate many more neutrons resulting
in extremely neutron-rich bound nuclei. Accommodation of more
neutrons in turn finally extends further the neutron drip-line. The
present study of isotones is intended to demonstrate the persistence
of this phenomenon in the case of neutron magic nuclei as well.
Thus, in the case of neutron magic isotones the proton drip-line,
instead of neutron drip-line, is found to be extended. However, in
contrast to the neutron drip-line case of proton magic nuclei, here
the effect of extending the proton drip-line is not seen to be
pronounced. This is due to the disruptive effect of Coulomb forces
which very much limit the number of protons being accommodated while
keeping the isotonic nucleus bound.

Moreover, as per the definition of two proton (neutron) drip-lines
the nuclei become unbound as soon as the two proton (neutron)
separation energy approaches zero. However, some nuclei even with
negative two proton separation energy may have finite life time due
to the combined effect of Coulomb and centrifugal barriers. This
life time may be large enough to allow even such a nucleus to be
studied experimentally. This effect in combination with  the
resonant states phenomenon described above effectively further
pushes the proton drip-line with nuclei having negative two proton
separation energy.

The shell closures with pronounced gaps between shells  in nuclei
endow them with spherical shape. Consequently, the magic nuclei are
characterized with zero  deformation. In our systematic
investigations we  first carry out RMF+BCS calculations including
the deformation degree of freedom (to be referred to throughout as
deformed RMF+BCS) to establish whether the entire chain of magic
isotones for a given neutron number is indeed spherical or not. It
is gratifying to note that leaving aside a few exceptions away from
the line of stability in the case of \textit{N} = 40 and also to
some extent for the neutron-rich \textit{N} = 28 isotones,
invariably the entire chain of nuclei shows negligible deformation,
especially for \textit{N} = 8, 20, 50, 82 and 126 where entire chain
shows zero deformation.

In the case of negligible/zero deformation, we take advantage of the
RMF+BCS approach for spherical shapes (to be referred to throughout
as spherical RMF+BCS) for the analysis of results in terms of
spherical single particle wave functions and energy levels to make
the discussion of shell closures and magicity more convenient and
transparent. Also, behavior of the single particle states near the
Fermi surface which in turn plays an important role near the
drip-line can be easily understood. Moreover,  within such a
framework contributions of neutron and proton single particle states
to the density profiles, pairing gaps, total pairing energy and
separation energy which are also equally important in the study of
exotic phenomena can be demonstrated with clarity. This approach
indeed turns out to be very useful for the study of poorly
understood exotic nuclei.

Our spherical RMF+BCS calculations have been performed with two
different force parametrization, TMA and NL-SH \cite{suga,sharma} to
check if the results have any dependence on force parametrization.
Details of our calculations show that the two interactions employed
here produce very similar results. Therefore, unless required, in
most of the cases we have presented the results obtained with the
TMA force \cite{suga} only.

\subsubsection{Proton-Rich Nucleus $^{58}_{30}$Zn$_{28}$ at the
drip-line:}

\begin{figure}[th]
\centerline{\psfig{file=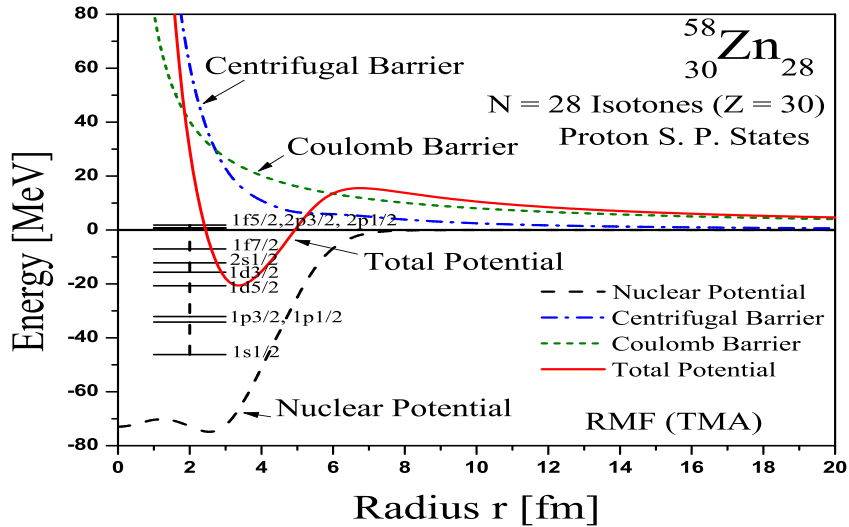,width=14cm,height=8cm}} \caption{The
RMF potential energy (sum of the scalar and vector potentials), for
the nucleus $^{58}_{30}$Zn$_{28}$ as a function of radius is shown
by the black long dashed line. The centrifugal barrier energy for
proton resonant states 1f$_{5/2}$ and coulomb barrier are also shown
in the figure by dot dashed blue line and dashed green line
respectively. The red solid line represents total potential, the sum
of RMF potential energy, the centrifugal barrier energy and the
coulomb barrier. The figure also shows the energy spectrum of the
proton single particle states lying close to the Fermi level and
include both the bound and continuum states. The important resonant
states 1f$_{5/2}$, 2p$_{3/2}$ and 2p$_{1/2}$ at 0.57, 0.84 and 1.82
MeV, respectively are also depicted in the figure.}
\end{figure}

In order to demonstrate our results for the proton-rich nucleus
$^{58}_{30}$Zn$_{28}$  of  the \textit{N} = 28 isotonic chain, we
have plotted in Fig. 1 the calculated RMF potential, a sum of scalar
and vector potentials, along with the spectrum of the bound proton
single particle states. Though this is a typical example of the
proton-rich case, the main features of the potential, single
particle spectrum and wave functions besides some finer details
remain valid for all the isotones. The figure also shows the
positive energy proton single particle states corresponding to the
low-lying resonances 1f$_{5/2}$, 2p$_{3/2}$ and 2p$_{1/2}$ close to
the Fermi surface. Amongst these, it is observed that especially the
resonant states  play significant role in the binding of proton-rich
isotone $^{58}_{30}$Zn$_{28}$ through their contributions to the
total pairing energy. In contrast to other states in the box, the
respective position of the resonant $1f_{5/2}$, 2p$_{3/2}$ and
2p$_{1/2}$ states is not much affected by changing the box radius
around $R = 30$ fm. For the purpose of illustration we have also
depicted in Fig. 1 centrifugal barrier of $1f_{5/2}$ states and
coulomb barrier. The total mean-field potential, obtained by adding
the centrifugal potential energy and coulomb barrier energy is also
shown. It is evident from the figure that the total effective
potential for the 1f$_{5/2}$ state including coulomb barrier, has an
appreciable barrier for the trapping of waves  to form a quasi-bound
or resonant state. Such a meta-stable state remains mainly confined
to the region of the potential well and the wave function exhibits
characteristics similar to that of a bound state.

\begin{figure}[th]
\centerline{\psfig{file=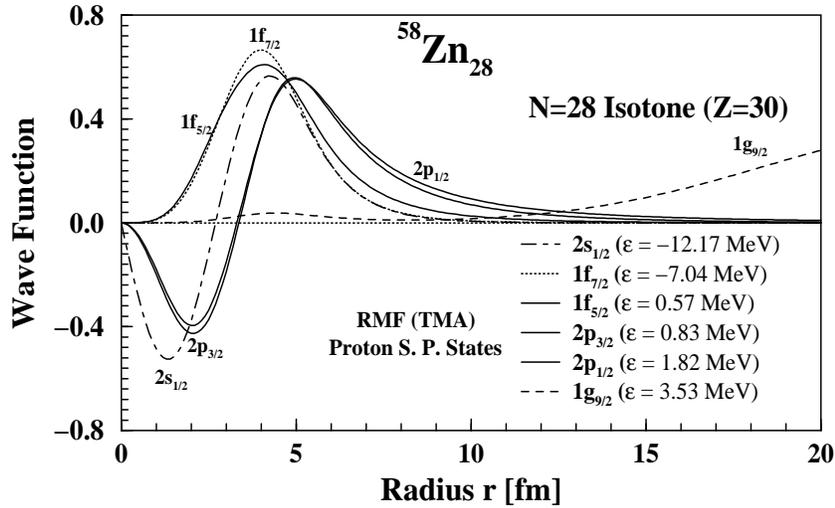,width=11cm,height=7cm}}
\caption{Radial wave functions of a few representative proton single
particle states with energy close to the Fermi surface for the
nucleus $^{58}_{30}$Zn$_{28}$. The resonant states 1f$_{5/2}$,
2p$_{3/2}$ and 2p$_{1/2}$ at energies 0.57, 0.84 and 1.82 MeV,
respectively, are shown by solid lines, while the bound 2s$_{1/2}$
state at -12.17 MeV, and 1f$_{7/2}$ state at -7.04 MeV are shown by
long dashed and dotted lines. The continuum 1g$_{9/2}$ state at 3.53
MeV has been depicted by small dashed line. It is seen that the
resonant states in the continuum have characteristics similar to the
bound states and are confined within the potential region.}
\end{figure}

This is clearly observed in Fig. 2 which depicts the radial wave
functions of some of the proton single particle states lying close
to the Fermi surface with the proton Fermi energy being $\lambda_p\,
=\,-0.097$ MeV. These include the bound $1f_{7/2}$ and $2s_{1/2}$
states, in addition to the resonant $1f_{5/2}$, $2p_{3/2}$ and
$2p_{1/2}$ states. As an example of a non-resonant state, the figure
also depicts a typical high lying continuum $1g_{9/2}$ state at 3.53
MeV. The wave functions for the  1f$_{5/2}$, 2p$_{3/2}$ and
2p$_{1/2}$ states in Fig. 2  are clearly seen to be confined within
a radial range of about 8 fm, and have a decaying component outside
this region characterizing resonant states. Such type of states thus
have a good overlap with the bound states near the Fermi surface
leading to significant contribution to the pairing gap value
$\Delta_{1f_{5/2}}$, $\Delta_{2p_{3/2}}$, $\Delta_{2p_{1/2}}$ and
also to the total pairing energy of the system near the
drip-line.\vspace{0.2cm}

In contrast, the main part of the wave function for the non-resonant
states, e.g. $1g_{9/2}$, is seen to be spread over outside the
potential region, though a small part is also contained inside  the
potential range. This type of state thus has a poorer overlap with
the bound states near the Fermi surface leading to small value for
the pairing gap $\Delta_{1g_{9/2}}$. Similarly, other positive
energy states lying away from the Fermi level, for example,
$1h_{11/2}$, $1i_{13/2}$ etc. have a negligible contribution to the
total pairing energy of the system. These features can be seen from
Fig. 3 which depicts the calculated pairing gap energy $\Delta_j$
for some of the proton states in the nucleus $^{58}_{30}$Zn$_{28}$.
However, we have not shown in the figure the single particle  states
having negligibly small $\Delta_j$ values as these do not contribute
significantly to the total pairing energy.

\begin{figure}[th]
\centerline{\psfig{file=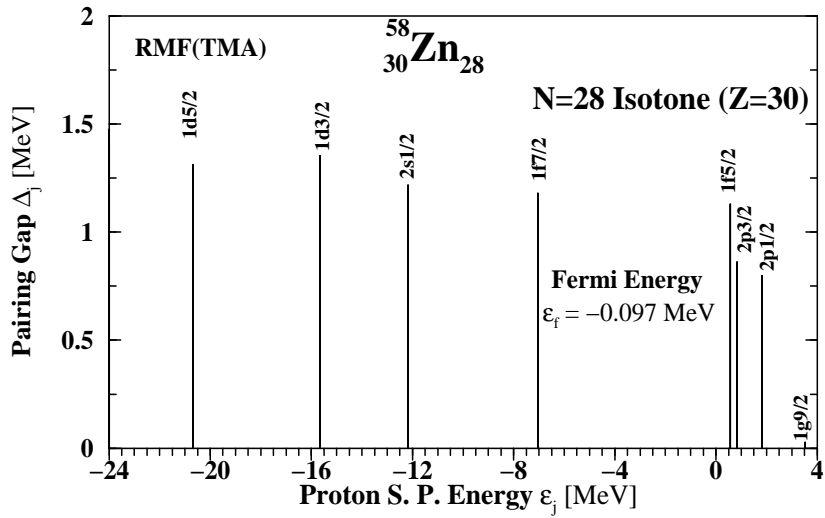,width=11cm,height=7cm}}
\caption{Pairing gap energy $\Delta_{j}$ of proton single particle
states with energy close to the Fermi surface for the nucleus
$^{58}_{30}$Zn$_{28}$. The resonant 1f$_{5/2}$, 2p$_{3/2}$ and
2p$_{1/2}$ states at energy 0.57, 0.84 and 1.82 MeV, respectively,
have the gap energy of about 1 MeV which is close to that of bound
states like 1f$_{7/2}$, 2s$_{1/2}$ etc.}
\end{figure}

One observes indeed in Fig. 3 that the gap energy for the $1f_{5/2}$
state has a value close to 1 MeV which is quantitatively similar to
that of bound states $1f_{7/2}$ and $2s_{1/2}$. Also, Fig. 3 shows
that the pairing gap value for the non-resonant states in the
continuum , like $1g_{9/2}$, have negligible gap energy close to
zero. On further addition of two protons to the drip-line nucleus
$^{58}_{30}$Zn$_{28}$, it yields the heavier isotone
$^{60}_{32}$Ge$_{28}$. This nucleus $^{60}_{32}$Ge$_{28}$ has
negative two proton separation energy and lies beyond the proton
drip-line and it is a good candidate of two proton decay as observed
in Ref.~\refcite{stolz}. This nucleus decays by emission of two
proton with the combined effect of the Coulomb and centrifugal
barrier as has been discussed elaborately in Ref.~\refcite{saxena}.
This is in contrast to the proton magic nuclei which are
neutron-rich nuclei and lie near the neutron drip-line, for example
the heavier isotopes of Ca \cite{yadav,yadav1}. As explained in
Ref.~\refcite{yadav1}, on further addition of neutrons  in the case
of neutron-rich Ca isotopes the single particle states like
$3s_{1/2}$, $1g_{9/2}$ and $2d_{5/2}$ which lie near the Fermi level
gradually come down close to zero energy on increasing no. of
neutrons. Even the $1g_{9/2}$ and $3s_{1/2}$ states become bound
states. This helps in accommodating more and more neutrons which are
just bound. In fact, the occupancy (no. of particle in the state) of
the $3s_{1/2}$ state in these extremely neutron-rich isotopes causes
the halo formation \cite{yadav1}. This difference between the
neutron-rich isotopes, and proton-rich isotones involving similar
single particle states is solely caused by the large repulsive
Coulomb interaction with further addition of protons. This is easily
demonstrated by comparing the situations, for example, in the
heavier Ca isotopes like $^{62-72}$Ca \cite{yadav1}, with that in
the heavier \textit{N}=28 isotones like $^{58}_{30}$Zn$_{28}$ and
$^{60}_{32}$Ge$_{28}$.

RMF calculations \cite{geng1} show that this nucleus
$^{60}_{32}$Ge$_{28}$ has two proton separation energy $S_{2p}$ =
-0.52 and one proton separation energy $S_{p}$ = 0.27 and may be a
possible candidate of two proton radioactivity \cite{stolz} due to
the combined barrier provided by the Coulomb and centrifugal
effects. The physical situation of proton-rich $^{60}_{32}$Ge$_{28}$
nucleus is similar to the previous nucleus $^{58}_{30}$Zn$_{28}$.
The states 1f$_{5/2}$, 2p$_{3/2}$ and 2p$_{1/2}$  at energy 0.74,
1.13 and 2.12 MeV, respectively are resonant states for this
nucleus. Such nuclei close to the drip-line or even beyond the
drip-line have been the focus of experimental studies during the
last few years employing the latest new technological advancements
in the field of radioactive ion beams. Such nuclei may be good
candidates of two proton radioactivity which is experimentally
verified recently in the decay of $^{45}$Fe
\cite{Pftzner,Giovinazzo}, and subsequently in other experiments in
the decay of $^{54}$Zn \cite{Blank,Ascher} and $^{48}$Ni
\cite{Dossat,Pomorski}. RMF description of such nuclei has been
presented elaborately in Ref.~\refcite{saxena}.

\subsubsection{Two proton separation energy, Single particle
energy:}

The contribution of pairing energy plays an important role for the
stability of the nuclei near the drip-lines and consequently in
deciding the position of the neutron and proton drip-lines. The
proton shell closure as indicated by the  calculated results for
\textit{Z} = 20 and 28 along with the neutron shell closure for the
isotones with \textit{N} = 28 render the nuclei $^{48}$Ca and
$^{56}$Ni doubly magic. However, it is found that the \textit{N} =
28 shell closure gets weakened in the proton deficient side and at
\textit{Z} = 12 for the nucleus $^{40}$Mg the shell closure
\textit{N} = 28 quenches which is in accordance with recent study of
disappearance of \textit{N} = 28 magicity near drip-line \cite{Li}.

\begin{figure}[th]
\centerline{\psfig{file=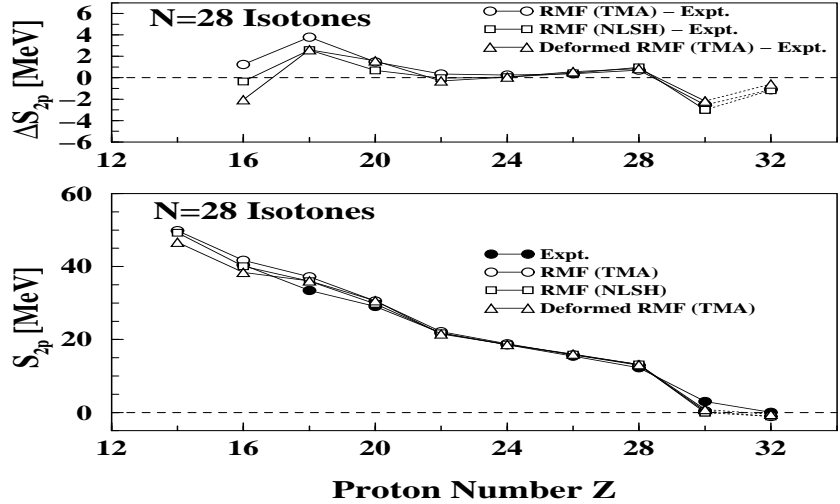,width=11cm,height=7cm}} \caption{Lower
Panel: Two proton separation energy S$_{2p}$ for the nuclei of
\textit{N} = 28 isotonic chain. Spherical RMF+BCS calculations
carried out with the TMA and NL-SH force parameters are compared
with the available experimental data and with the deformed RMF+BCS
calculations using the TMA force parametrization. Upper Panel: The
difference $\Delta S_{2p}$ of the calculated results for the two
proton separation energy with respect to that of the available
experimental data.}
\end{figure}

The shell structure as revealed by the pairing energies are also
exhibited in the variation of two proton separation energies
S$_{2p}$ as shown in the lower panel of Fig. 4 for the \textit{N} =
28 isotones. It is observed from the lower panel of the figure that
the spherical RMF+BCS calculations employing the TMA and NL-SH force
parametrization are similar in nature, which in turn are also seen
to be close to the deformed RMF+BCS calculations. A comparison of
these three calculations with the available experimental data for
the two proton separation energy shows that these are in fairly good
agreement with the measurements. However, For the purpose of a
closer comparison, we have  explicitly displayed in the upper panel
of Fig. 4 the differences $\Delta S_{2p}$ between the calculated
results and the experimental data for the two-proton separation
energy. It is observed that the absolute value of the  difference as
expressed by $\Delta S_{2p}$ varies from a minimum of close to zero,
to a maximum of about 4 MeV for the entire chain of \textit{N} = 28
isotonic nuclei.

Furthermore, as expected, it is observed from the lower panel of
Fig. 4 that an abrupt decrease in the  two proton separation energy
S$_{2p}$, occurs for the isotone lying next to a magic number. Once
again this  supports robust shell closures at \textit{Z} = 20 and 28
as can be seen from the lower panel of Fig. 4. Also, from this
figure it is clearly seen that for the spherical as well as for the
deformed RMF+BCS calculations the two proton drip-line lies at
\textit{Z} = 30. The S$_{2p}$ value at the drip-line for the
spherical as well as deformed RMF+BCS calculation using TMA force
parameters is found to be positive. Beyond \textit{Z} = 30, the
S$_{2p}$ value becomes negative as can be seen from the lower panel
of Fig. 4.

\begin{figure}[th]
\centerline{\psfig{file=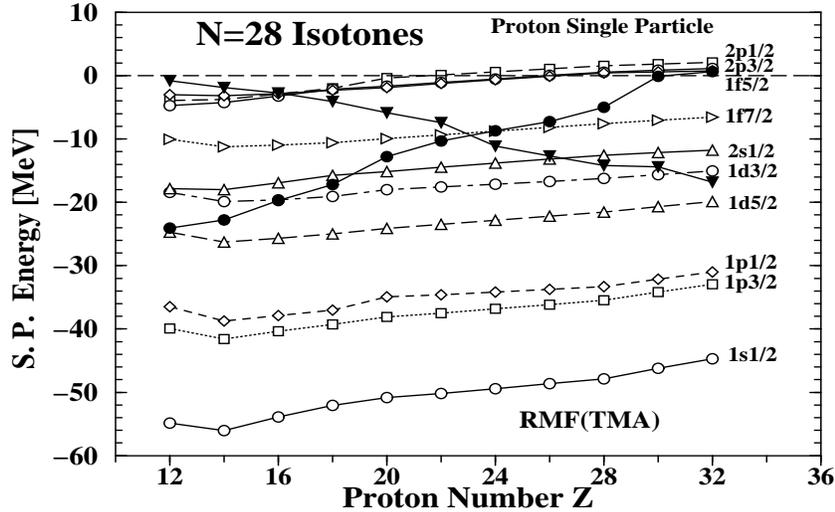,width=11cm,height=7cm}} \caption{
Variation of the proton single particle energies obtained in the
spherical RMF+BCS calculations with the TMA force for the \textit{N}
= 28 isotones with increasing proton number. The proton Fermi level
has been shown by filled circles connected by solid line, while the
neutron Fermi level is depicted by filled inverted triangles also
connected by solid line to guide the eyes.}
\end{figure}

In the spherical cases the structure of proton (or neutron) single
particle energy and its variation with increasing proton (or
neutron) number for the nuclei in general plays an important role in
the understanding and explanation of position of drip-lines,
variation in the proton or neutron radii and density distributions.
Thus for the isotone $^{56}_{28}\rm Ni_{28}$  the proton and neutron
single particle structures indicate shell closures for both proton
and neutron due to a large gap  between the single particle state
1f$_{7/2}$ and the next high lying states 1f$_{5/2}$ and 2p$_{3/2}$.
In the case of proton single particle states this gap is of about 8
MeV as can be seen in Fig. 5. Similar gap, though slightly less
pronounced, between the proton single particle state 2s$_{1/2}$ and
the next high lying 1f$_{7/2}$ state gives rise to shell closures at
proton number \textit{Z} = 20 as is clearly seen in Fig. 5. Thus,
the isotone $^{48}_{20}\rm Ca_{28}$ is seen to be a doubly magic
nucleus. A large energy gap is also seen between the 1d$_{5/2}$ and
1d$_{3/2}$ proton single particle states and the proton number
\textit{Z} = 14 is expected to produce shell closure.

From Fig. 5 it is observed that beyond the proton number \textit{Z}
= 28, the proton single particle states 1f$_{5/2}$ and 2p$_{3/2}$
lie in the continuum having energy 0.57 MeV  and 0.84 MeV,
respectively. The proton Fermi energy shown by the filled circles
connected by the solid line also moves up to be very close to zero
energy value ($\epsilon_{f} = $ -0.097 MeV). Consequently, further
addition of 2 protons beyond \textit{Z} = 28, partially fills the
positive energy proton single particle states 1f$_{5/2}$ and
2p$_{3/2}$ which act as resonant states for the nucleus
$^{58}_{30}$Zn$_{28}$ as has been described earlier in sec. 3.1.1.
The occupancy i.e. number of particles occupying the levels, of
these states with increasing proton number has been shown in Fig. 6.

These resonant states are akin to the bound states having their wave
functions confined within the nuclear potential region, and this
helps to make the isotone $^{58}_{30}$Zn$_{28}$ a bound nucleus
located close to the proton drip-line. Another addition of 2 protons
gives the isotone $^{60}_{32}$Ge$_{28}$ wherein the occupancy of the
proton single particle states 1f$_{5/2}$ and 2p$_{3/2}$ increases
further as is seen in Fig. 6. However, beyond \textit{Z} = 30, the
proton drip-line is reached the next isotone $^{60}_{32}$Ge$_{28}$
is found to be unstable against 2 proton emission as has been
discussed earlier.

\begin{figure}[th]
\centerline{\psfig{file=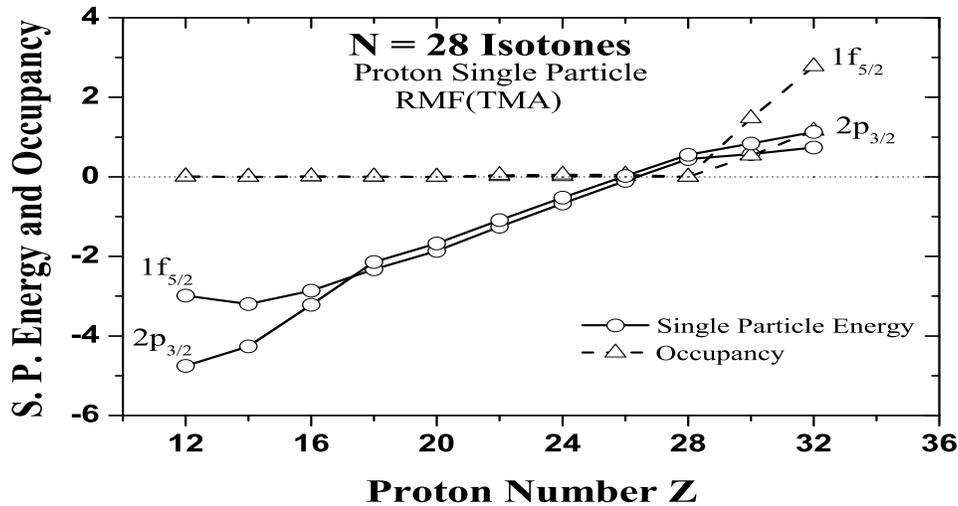,width=14cm,height=8cm}}
\caption{Variation as a function of increasing proton number
\textit{Z} in energy shown by open circles, and occupancy (number of
particles occupying the levels) depicted by open triangles for the
proton single particle states 1f$_{5/2}$ and 2p$_{3/2}$ which act as
resonant states in the proton-rich nuclei (\textit{Z} $>$ 28) of the
\textit{N} = 28 isotonic chain.}
\end{figure}

As remarked in previous section, even though such a nucleus is
unbound its decay by 2 proton emission is appreciably delayed due to
the barrier produced by the Coulomb interaction. In fact the
resonant proton single particle states 1f$_{5/2}$ and 2p$_{3/2}$
could accommodate all together 10 particles beyond \textit{Z} = 28
and thus would have in effect pushed the proton drip-line far away
to a large proton number \textit{Z} instead of \textit{Z} = 30 in
analogy, for example,  to the case of Ca isotopes wherein the low
lying resonant states 1g$_{9/2}$ accommodates 10 neutrons to enable
the existence of highly neutron-rich heavy  Ca isotopes
\cite{yadav,yadav1}. However,  due to the disruptive Coulomb
interaction amongst protons, the existence of heavier \textit{N} =
28 isotones beyond \textit{Z} =30, that is heavier than the isotone
$^{60}_{32}$Ge$_{28}$, with a life time large enough to enable
experimental measurements  as in the case of unbound
$^{60}_{32}$Ge$_{28}$ isotone becomes rather difficult.

\subsection{Neutron Magic Nuclei of \textit{N} = 8, 20, 40, 50, 82 and 126 Isotonic Chains}

In the present section it is shown that the neutron magic isotones
with neutron number \textit{N} = 8, 20, 50, 82 and 126, as well as
the submagic isotones corresponding to \textit{N} = 40 exhibit many
characteristics in common to each other and also to that of
\textit{N} = 28 isotones. Thus, in order to save space, here we have
described collectively together, though in brief, the results for
the nuclei forming the isotonic chains  with neutron magic and
submagic numbers. Again, instead of displaying the properties of
nuclei in each isotonic chain separately, to save space we have
considered only a specific isotone, especially the proton-rich one,
from each chain for the purpose of illustration.

The deformed RMF+BCS calculations indicate that all the isotones
with neutron number \textit{N} = 8, 20, 50, 82 and 126 are almost
spherical with zero quadrupole deformation parameter. Thus the
chains of these isotones can be conveniently described within the
spherical approach. In contrast, as expected, in the case of
submagic \textit{N} = 40 isotonic chain  the proton-rich isotones
for \textit{Z} $\geq$ 32 are well deformed. Thus only nuclei with
the spherical shape of \textit{N} = 40 isotonic chain have been
included in the spherical RMF+BCS description along with the
isotones of the chains with \textit{N} = 8, 20, 50, 82 and 126.

Similar to the case of heavy proton-rich nuclei in the  \textit{N} =
28 isotonic chain, the proton-rich nuclei lying close to the proton
drip-line in the isotonic chain of neutron magic numbers \textit{N}
= 40 and 126, and to a lesser extent those in the isotonic chains of
\textit{N} = 50 and 82, exhibit characteristics of having proton
resonant states which help in accommodating more protons giving rise
to the existence of  exotic  proton-rich nuclei. As opposed to the
case of isotones with \textit{N} = 28, 50, 82, 126 and 40, low lying
proton resonant states are not found in nuclei of the isotonic
chains with \textit{N} = 8 and 20.

With this in view, we have chosen the nuclei $^{46}_{26}$Fe$_{20}$,
$^{96}_{46}$Pd$_{50}$, $^{154}_{72}$Hf$_{82}$ and
$^{220}_{94}$Pu$_{126}$  as the representative examples of the
proton-rich nuclei in the  \textit{N} = 20, 50, 82 and 126 neutron
magic isotonic chains and describe their features in detail obtained
within the spherical RMF+BCS approach. This  aims to elucidate the
typical features of the calculated potential, single particle wave
function and the pairing gap energy etc. We have not included here
the description of any proton-rich isotone for \textit{N} = 40 since
these are well deformed as mentioned above. Also, in order to save
space, we have not included any proton-rich isotone with \textit{N}
= 8 as this case is very similar to that of the proton-rich
\textit{N} = 20 isotones.

In Fig. 7 we have plotted using solid lines the RMF potentials, a
sum of scalar and vector potentials, for the representative
proton-rich $^{46}_{26}$Fe$_{20}$, $^{96}_{46}$Pd$_{50}$,
$^{154}_{72}$Hf$_{82}$ and $^{220}_{94}$Pu$_{126}$ nuclei mentioned
above along with the spectrum for the bound proton single particle
states.\vspace{ 0.1cm}

\begin{figure}[th]
\centerline{\psfig{file=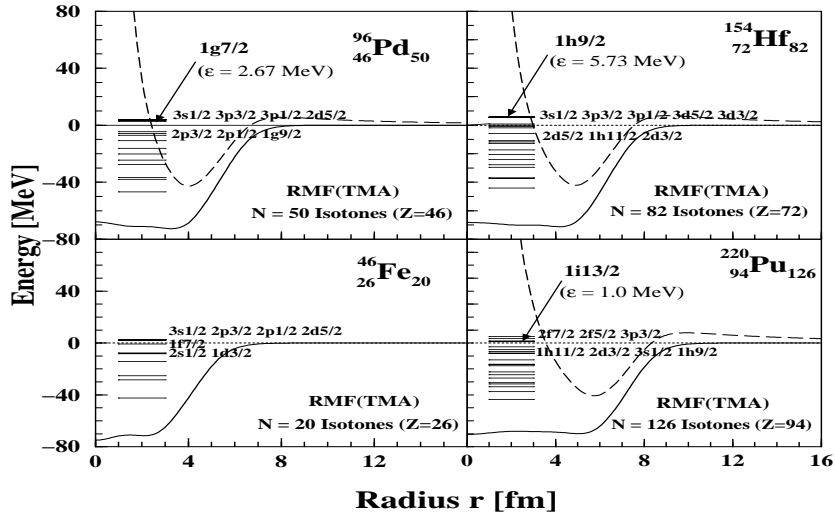,width=11cm,height=7cm}} \caption{The
RMF potential energy plots (sum of the scalar and vector potentials)
for the proton-rich $^{46}_{26}$Fe$_{20}$, $^{96}_{46}$Pd$_{50}$,
$^{154}_{72}$Hf$_{82}$ and $^{220}_{94}$Pu$_{126}$ nuclei of the
isotonic chain with neutron number \textit{N} = 20, 50, 82 and 126,
respectively, as a function of radius. Besides the position of low
lying resonant states, the plots also show the energy spectrum of
the bound and continuum proton single particle states. The long
dashed line represents the sum of RMF potential energy and the
centrifugal barrier energy for the proton resonant states mentioned
in the plots. No low lying resonant state occurs in the case of
\textit{N} = 20 proton-rich isotones such as $^{46}_{26}$Fe$_{20}$.}
\end{figure}

\noindent The plots also depict a few positive energy proton single
particle states near the Fermi level, besides the low lying resonant
states which contribute to the pairing energy. These resonant states
are respectively, $1g_{7/2}$ at energy 2.67 MeV in
$^{96}_{46}$Pd$_{50}$, $1h_{9/2}$ at energy 5.73 MeV in
$^{154}_{72}$Hf$_{82}$ and $1i_{13/2}$ at energy 1.00 MeV in
$^{220}_{94}$Pu$_{126}$. For these resonant states we have also
shown by long dashed lines the total mean-field potential given by a
sum of RMF potential energy and centrifugal barrier energy. It is
evident from  Fig. 7 that the effective total potential for the
resonant states have an appreciable barrier for the confinement of
waves to form a quasi-bound state.

\begin{figure}[th]
\centerline{\psfig{file=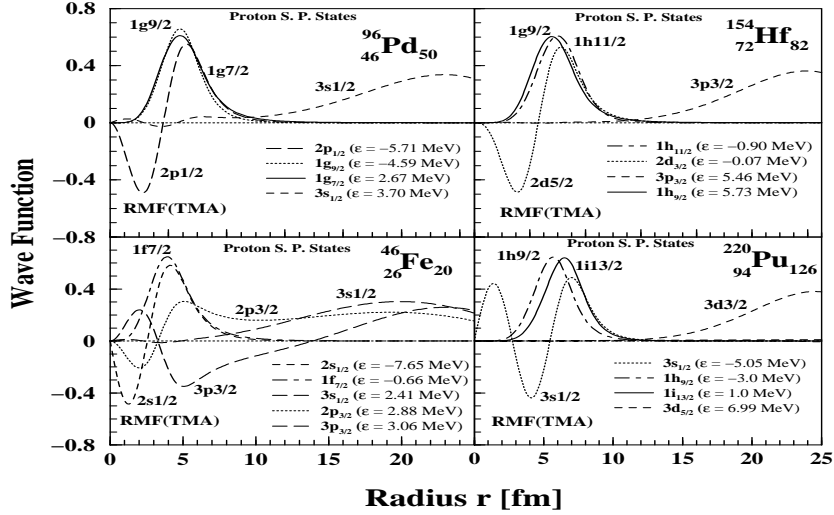,width=11cm,height=7cm}}
\caption{Radial wave functions of a few representative proton single
particle states with energy close to the Fermi surface for the
proton-rich $^{46}_{26}$Fe$_{20}$, $^{96}_{46}$Pd$_{50}$,
$^{154}_{72}$Hf$_{82}$ and $^{220}_{94}$Pu$_{126}$ nuclei belonging
to the  isotonic chains of \textit{N} = 20, 50, 82 and 126,
respectively, are shown. The proton resonant states for different
nuclei have been depicted by solid lines. No low lying resonant
state is found in nuclei belonging to \textit{N} = 20 and \textit{N}
= 8 isotonic chains.}
\end{figure}

It is seen from Fig. 7 that with increasing proton number, as
expected, the density of states increases. Moreover, with the
increasing number of neutrons and protons the total RMF potential
region also gets appreciably larger.  Also, as described in the case
of proton-rich  \textit{N} = 28 isotonic chain nucleus
$^{58}_{30}$Zn$_{28}$, it is evident from the plots that the
effective total potential for the states $1g_{7/2}$ in
$^{96}_{46}$Pd$_{50}$, $1h_{9/2}$ in $^{154}_{72}$Hf$_{82}$ and
$1i_{13/2}$ in $^{220}_{94}$Pu$_{126}$ has an appreciable barrier
for the confinement of waves to form a quasi-bound or resonant
state. The wave functions of these states exhibit characteristics
similar to that of a bound state as is seen in the plots shown in
Fig. 7. In contrast, the main part of the wave functions for the
non-resonant states for these nuclei is seen to be mostly spread
over outside their potential region. The wave function of the proton
single particle states $3s_{1/2}$ in $^{46}_{26}$Fe$_{20}$ and
$^{96}_{46}$Pd$_{50}$, $3p_{3/2}$ in $^{154}_{72}$Hf$_{82}$, and
$3d_{3/2}$ in $^{220}_{94}$Pu$_{126}$ displayed in Fig. 8 represent
typical examples of such states. No low lying resonant states are
found in the case of proton-rich nuclei of the isotonic chains with
\textit{N} = 20 and \textit{N} = 8. For the nucleus
$^{46}_{26}$Fe$_{20}$ illustrated here as an example of proton-rich
\textit{N} = 20 isotone, the possible proton resonant state
$1f_{5/2}$ lies much above the proton Fermi level in the continuum.
Similar situation prevails for the nuclei belonging to  \textit{N} =
8 isotonic chain. In both cases the states in the continuum close to
the proton fermi level are low angular momentum states, viz.
$3s_{1/2}$, $2p_{3/2}$, $2p_{1/2}$, $2d_{5/2}$ etc. The effective
total potential for these states does not seem to have appreciable
barrier to form a quasi-bound state as is clearly seen from the wave
functions of some of these states displayed in Fig. 8. The above
mentioned characteristics of resonant and non-resonant states are
also seen in their contributions to the pairing gap energies as has
been described below.

The wave functions for the resonant states are clearly seen to be
confined within a radial range of about 10 fm, and have a decaying
component outside this region. Such type of states may have a good
overlap with the bound states near the Fermi surface leading to
significant contribution to the respective pairing gap values, and
also to the total pairing energy of the system near the drip-line.

In Fig. 9 we have displayed the pairing gap energy values of proton
single particle states located close to the Fermi level. It is seen
from Fig. 9 that the pairing gaps  for the single particle proton
states in the representative nucleus $^{46}_{26}$Fe$_{20}$ for the
isotonic chain \textit{N} = 20 near the Fermi level range
approximately between 0.9 MeV to 1.3 MeV, where the Fermi energy
lies at $\epsilon_f$ = -0.041 MeV. In contrast to other
representative cases shown in Fig. 9, viz. $^{96}_{46}$Pd$_{50}$,
$^{154}_{72}$Hf$_{82}$ and $^{220}_{94}$Pu$_{126}$, it is seen that
only the lowest proton single particle states are occupied according
to the number of protons, and the pairing interaction does not
couple the bound states near the Fermi level with the positive
energy states. Thus in the case of proton-rich $^{46}_{26}$Fe$_{20}$
nucleus the positive energy states are not populated. A similar
situation is seen to prevail in the case of nuclei belonging to the
isotonic chain with \textit{N} = 8.

The next two plots in Fig. 9 show the pairing gap energies for the
nuclei $_{46}^{96}$Pd$_{50}$ and $_{72}^{154}$Hf$_{82}$ representing
the proton-rich examples of the isotonic chain with \textit{N} = 50
and \textit{N} = 82 respectively. Again, it is clearly seen in the
figure that the resonant  proton single particle states 1g$_{7/2}$
($\epsilon$ = 2.67 MeV) in $_{46}^{96}$Pd$_{50}$ and 1h$_{9/2}$
($\epsilon$ = 5.73 MeV) in $_{72}^{154}$Hf$_{82}$  have large
pairing gap values similar to that of bound proton single particle
states. These resonant states lie far away from the proton Fermi
level which is located at  $\epsilon_f$ = -4.11 MeV in
$_{46}^{96}$Pd$_{50}$ and at $\epsilon_f$ = -0.487 MeV in
$_{72}^{154}$Hf$_{82}$. Due to such energy difference the resonant
states in these nuclei are not able to connect to the states near
the Fermi level through the pairing correlations and, therefore,
their contribution to the pairing energy is also rather small. This
also implies that the effect of resonant states in accommodating
more protons to the system near the proton drip-line, and hence
pushing the drip-line further is not prominent in the case of
$^{96}_{46}$Pd$_{50}$ and $^{154}_{72}$Hf$_{82}$ nuclei of the
isotonic chains with \textit{N} = 50 and \textit{N} = 82.

For the nucleus $_{94}^{220}$Pu$_{126}$ representing the isotonic
chain with neutron number \textit{N} = 126 the proton single
particle resonant state is 1i$_{13/2}$ as shown in Fig. 9. This
resonant state lies at energy $\epsilon$ = 1.00 MeV, whereas the
proton Fermi level is located at $\epsilon_f$ = -0.149 MeV. Again,
it is seen that the pairing gap energy of the resonant 1i$_{13/2}$
state has a value $\Delta_{j}\approx 1.2$ MeV, which is close to
that of the bound states.\vspace{ 0.5cm}

\begin{figure}[th]
\centerline{\psfig{file=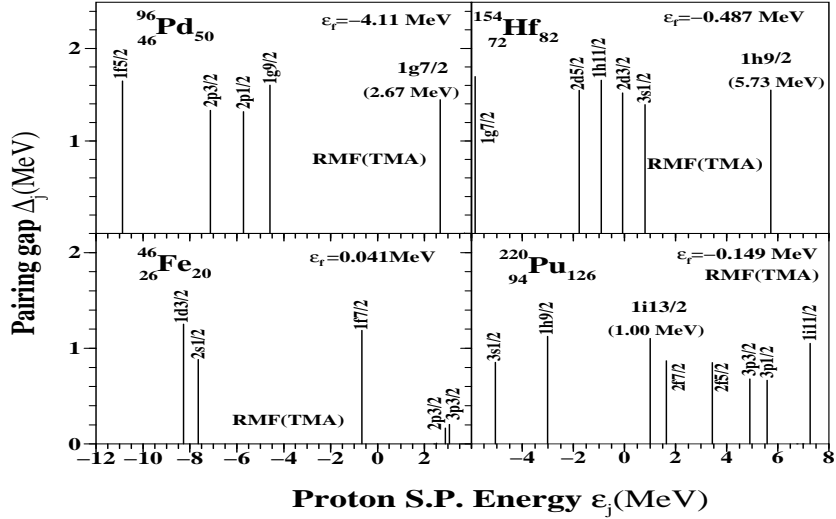,width=11cm,height=7cm}}
\caption{Pairing gap energy $\Delta_{j}$ of proton single particle
states with energy close to Fermi level for the proton-rich nuclei
$^{46}_{26}$Fe$_{20}$, $^{96}_{46}$Pd$_{50}$, $^{154}_{72}$Hf$_{82}$
and $^{220}_{94}$Pu$_{126}$ representing the isotonic chains with
\textit{N} = 20, 50, 82 and 126, respectively.}
\end{figure}

As can be seen in the plot for the $_{94}^{220}$Pu$_{126}$ isotone
in Fig. 9, there are other continuum  states, viz. 2f$_{7/2}$,
2f$_{7/2}$, 3p$_{3/2}$, 3p$_{3/2}$ and 1i$_{11/2}$ which lie at
higher energy and have relatively appreciable values for the pairing
gap energy.  The proton 1i$_{13/2}$  state being a resonant state
has good overlap with the bound states and, therefore, is found to
have sizeable contribution to the pairing energy. This  helps in the
occurrence of bound proton-rich isotone $_{94}^{220}$Pu$_{126}$
located at the two proton drip-line for the \textit{N} = 126
isotonic chain. Other  states  mentioned above; though these also
have appreciable pairing gap energy, are found to have no
substantial overlap with the bound states lying far down in energy.
This implies that the occupancy of these continuum states is
negligible and their contribution to the total pairing energy is not
substantial. Contrary to this, the resonant single particle state
1i$_{13/2}$  is found to be filled in by $\approx$ 2 protons.

\section{Summary}
In the present investigations we have employed relativistic
mean-field plus BCS (RMF + BCS) approach
\cite{walecka,serot,pgr,pgr1,gambhir,gambhir1,suga,ring,yadav,saxena,saxena1,geng1}
to carry out a systematic study for the ground state properties of
the entire chains of even-even neutron magic nuclei represented by
isotones of traditional neutron magic numbers \textit{N} = 8, 20,
28, 50, 82 and 126 as well as isotones of \textit{N} = 40,
considered to be neutron sub-magic. From the extensive calculations
it is established that the majority of isotones belonging to these
chains are indeed spherical. For this purpose we have employed the
deformed RMF+BCS approach wherein for the sake of simplicity only
axially deformed shapes are considered\cite{gambhir,gambhir1,ring}.
For our calculations in both approaches, the deformed RMF+BCS and
the spherical RMF+BCS, we have used the TMA \cite{suga} Lagrangian
density extensively used in the relativistic mean-field
calculations\cite{yadav,yadav1,geng1}. Further, in order to check
the validity of our description for different RMF force
parameterizations, we have carried out the spherical RMF+BCS
calculations using also the NL-SH \cite{ring,sharma} Lagrangian
density which has been equally popular for the relativistic
mean-field calculations.

One of the prime reason of this study has been to look into the role
of low lying resonant states which have been found earlier in the
investigations of proton magic isotopes of nuclei\cite{yadav,yadav1}
to act akin to the bound states leading to accumulation of
additional loosely bound neutrons. Eventually this resulted in the
existence of highly neutron-rich nuclei. In some cases, even the
occurrence of halo formation, for example in the heavy Ca and Zr
isotopes, has been predicted\cite{yadav,yadav1}. It is found that
the same mechanism does persist in these neutron magic nuclei and
the proton single particle resonant states play the same role here
and give extra stability to drip-line nucleus (nuclei). However, the
phenomenon gets restricted to the accumulation of only a few protons
due to the disruptive Coulomb forces amongst protons.

As an important result of the present studies we have described the
effect of proton single particle resonant states in accommodating
additional protons, which results effectively in extending the two
proton drip-line. This has been illustrated through the example of
proton-rich isotone $^{58}_{30}$Zn$_{28}$ lying at the two proton
drip-line of the \textit{N} = 28 isotonic chain. Similar to the case
of heavy proton-rich nuclei in the  \textit{N} = 28 isotonic chain,
the proton-rich nuclei lying close to the proton drip-line in the
isotonic chain of neutron magic numbers \textit{N} = 40 and 126, and
to a lesser extent those in the isotonic chains  of \textit{N} = 50
and 82, exhibit characteristics of having proton resonant states
which help in accommodating more protons giving rise to the
existence of exotic proton-rich nuclei. Towards this end, we have
chosen for the purpose of illustration the isotones
$^{46}_{26}$Fe$_{20}$, $^{96}_{46}$Pd$_{50}$, $^{154}_{72}$Hf$_{82}$
and $^{220}_{94}$Pu$_{126}$  as the representative examples of the
proton-rich nuclei in the  \textit{N} = 20, 50, 82 and 126 neutron
magic isotonic chains.

It is found for all the isotonic chains considered in the present
study that the calculated binding energy values obtained by
employing the deformed RMF+BCS approach are in very good agreement
with the available experimental data \cite{audi}. Since majority of
the isotones described here are spherical, these results are
obviously also in accord with the spherical RMF+BCS calculations as
in the case of the \textit{N} = 28 isotonic chain. Moreover, results
of our spherical RMF+BCS  calculations using TMA and NL-SH force
parameters produce similar results. This similarity in results is
seen to hold true for all the other physical quantities, viz. the
rms radii and densities for proton and neutron distributions etc.
indicating the force independence of these results.

\section*{Acknowledgements}
Support through a grant (SR/S2/HEP-01/2004) by the Department of
Science and Technology (DST), India, is acknowledged. The authors
are indebted to Dr. L. S. Geng, RCNP, Osaka University, Osaka,
Japan, for valuable correspondence.

\end{document}